\documentclass[12pt,preprint]{aastex}
\usepackage{epsf}
\usepackage{natbib}

\tabletypesize{\scriptsize}

\def\gappeq{\mathrel{ \rlap{\raise.5ex\hbox{$>$}}
                      {\lower.5ex\hbox{$\sim$}}  } }

\begin{document}
%\slugcomment{Accepted for Publication in ApJ}
\shorttitle{Exozodi Light Curves}
\shortauthors{Stark}

\title{The Transit Light Curve of an Exozodiacal Dust Cloud}

\author{Christopher C. Stark\altaffilmark{1}}

\altaffiltext{1}{Department of Terrestrial Magnetism, Carnegie Institution of Washington, 5241 Broad Branch Road, NW, Washington, DC 20015-1305;
cstark@dtm.ciw.edu}

\begin{abstract}

Planets embedded within debris disks gravitationally perturb nearby dust and can create clumpy, azimuthally asymmetric circumstellar ring structures that rotate in lock with the planet.  The Earth creates one such structure in the solar zodiacal dust cloud.  In an edge-on system, the dust ``clumps" periodically pass in front of the star as the planet orbits, occulting and forward-scattering starlight.  In this paper, we predict the shape and magnitude of the corresponding transit signal.  To do so, we model the dust distributions of collisional, steady-state exozodiacal clouds perturbed by planetary companions.  We examine disks with dusty ring structures formed by the planet's resonant trapping of in-spiraling dust for a range of planet masses and semi-major axes, dust properties, and disk masses.  We synthesize edge-on images of these models and calculate the transit signatures of the resonant ring structures.  The transit light curves created by dusty resonant ring structures typically exhibit two broad transit minima that lead and trail the planetary transit.  We find that Jupiter-mass planets embedded within disks hundreds of times denser than our zodiacal cloud can create resonant ring structures with transit depths up to $\sim10^{-4}$, possibly detectable with \emph{Kepler}.   Resonant rings produced by planets more or less massive than Jupiter produce smaller transit depths.  Observations of these transit signals may provide upper limits on the degree of asymmetry in exozodiacal clouds.

\end{abstract}

\keywords{circumstellar matter --- interplanetary medium --- planet--disk interactions --- methods: numerical --- planetary systems}

\section{Introduction}

The \emph{Kepler Mission} is searching for Earth-size planets in edge-on extrasolar planetary systems by detecting small photometric variations as planets transit in front of their host stars \citep{bkb10}.  To achieve this goal, \emph{Kepler} was designed to have a photometric precision of $\sim2 \times 10^{-5}$ over 6.5 hour timescales for $V=12$ Sun-like stars \citep{jcc10, kbb10}.  \emph{Kepler}'s unsurpassed photometric precision for exoplanet transits has enabled a wealth of other observations, including asteroseismological studies \citep{gjb10}, ellipsoidal star variations \citep{wos10}, and Doppler beaming binaries \citep{bmo11}.  \emph{Kepler} may also detect photometric variations caused by large satellites orbiting transiting planets \citep{ss99} as well as large, dense planetary rings \citep{bf04}.

Recent resolved images of the outer regions of several debris disks hint at another source of photometric variability: debris disk structures.  These resolved images reveal asymmetric structures in the form of rings, gaps, and clumps, likely created by gravitational perturbations from unseen extrasolar planets \citep[e.g.][]{g98, kgc05, g06, s09}.  Asymmetric structures have been observed in edge-on disks like $\beta$ Pictoris \citep{h00, g06} and AU Microscopii \citep{kag05}, and others have been observed to slowly rotate on timescales consistent with Keplerian orbital motion \citep{ghw05}.  Several massive hot disks have also been detected in the inner regions of nearby stars, where the orbital period is much shorter \citep[e.g.][]{szw05, adm06, b06}.  Could short-period asymmetries in the inner regions of edge-on disks produce photometric variations detectable with \emph{Kepler}?

Models of disk-planet interactions predict that the inner regions of debris disks may contain asymmetries \citep{sk08}.  As dust grains spiral inward via Poynting-Robertson (PR) drag, planets gravitationally perturb their orbits and can trap the dust into exterior mean motion resonances (MMRs).  This trapping can produce large-scale structure in the disk in the form of an overdense, azimuthally asymmetric circumstellar ring that orbits in lock with the perturbing planet.

At least one such structure exists in the inner regions of our own debris disk, the zodiacal cloud.  Infrared observations of the zodiacal cloud obtained with the DIRBE instrument onboard COBE confirmed excess flux leading and trailing the Earth's orbit \citep{r95}, interpreted as a circumsolar ring of dust resonantly trapped by the Earth \citep{jz89, d94}.  The inner regions of other debris disks, possibly much more dense than the zodiacal cloud, may harbor similar structures.

A resonant ring structure created by a planet on a circular orbit in an exozodiacal cloud typically exhibits the following asymmetries \citep{sk08}:
\begin{enumerate}
	\item A density deficit, or ``gap," near the location of the planet
	\item A density enhancement, or ``clump," trailing the planet
	\item A second clump leading the planet, typically with density less than the trailing clump
\end{enumerate}
The location of the leading and trailing clumps vary with parameters like planet mass and dust size, but for a planet with mass greater than a few Earth masses, these clumps typically lead and trail the planet by $\sim90^{\circ}$ in its orbit \citep{sk08}.  

For an edge-on system, these density enhancements would produce a variable photometric signal as the structure rotates in lock with the planet's orbit.  As a clump transits, the disk blocks additional starlight, producing a stellar extinction light curve with amplitude $\sim \Delta \tau_r$, where $\Delta \tau_r$ is the change in the radial optical depth along the line of sight to the star.  This amplitude can be approximated as
\begin{equation}
	\Delta \tau_r \sim \Delta \tau \; \frac{R}{2H},
\end{equation}
where $\tau$ is the face-on optical depth, $R$ is the radial width of the resonant ring structure, and $H$ is the scale height of the resonant ring.  For a disk with a range of inclinations on the order of $10^{\circ}$, $R/2H$ is on the order of a few for a high-contrast resonant ring structure \citep{sk08}.  However, \citet{sk08} showed that $\Delta \tau$ can approach $10 \tau_{\rm b}$ in the absence of grain-grain collisions, where $\tau_{\rm b}$ is the background optical depth, i.e. the optical depth interior or exterior to the resonant ring structure.  Exozodiacal clouds hundreds to thousands of times more dense than our zodiacal cloud, which has an optical depth $\tau \sim 10^{-7}$ near 1 AU, may be capable of producing a signal detectable with \emph{Kepler}.

However, another effect works against this signal.  The projected angular size of a dust clump is typically much larger than the angular size of the star.  During the transit of a dust clump the majority of the clump, which is near the line of sight to the star but not occulting the star, forward scatters starlight toward the observer; forward scattering increases the disk flux during dust clump transits.  Although this effect is likely small compared to the occultation of starlight, a robust calculation of an exozodiacal cloud's light curve must also include the disk's scattered light.

In this paper, we model the transit light curves of collisional exozodiacal clouds with asymmetric structures.  We limit our investigation to models of ring structures created by resonantly trapped in-spiraling dust.  We do not model other scenarios that may lead to structure, including resonant trapping of planetesimals and transient collisional phenomena, but briefly discuss these alternative scenarios in Section \ref{disklifetimes_section}.  We show sample light curves in Section \ref{results_section} and determine the disk and planet parameters that would lead to a maximum signal amplitude.  In Section \ref{discussion_section} we discuss whether such disks could exist and whether we could currently detect their transit light curves.

\section{Numerical Method \label{numericalmethod_section}}

We modeled the light curves for structured exozodiacal clouds using a two-step process.  First, we modeled the 3D dust distributions for steady-state collisional debris disks perturbed by single planet orbiting a Sun-like star.  Second, we illuminated these models with starlight and synthesized edge-on images as a function of planetary phase.

\subsection{Exozodi Models}
\label{exozodimodels}

For a signal $\sim 10^{-4}$, possibly detectable by \emph{Kepler}, the disk's background optical depth $\tau_{\rm b} \sim 10^{-5}$, i.e. hundreds of times greater than the zodiacal cloud.  In such dense disks the average time for two dust grains to collide, $t_{\rm coll}$, can be shorter than the Poynting-Robertson time, $t_{\rm PR}$, for the dust grains that dominate the disk's optical depth.  Modeling the light curve for these disks therefore requires a model that treats both resonant gravitational dynamics as well as grain-grain collisions.

We produced models of collisional exozodiacal clouds with resonant ring structures using a collisional grooming algorithm.  This algorithm simultaneously solves the equations of motion governing small dust grains in a planetary system and the number flux equation including destruction of dust grains via grain-grain collisions.  In short, the collisional grooming algorithm takes a collisionless debris disk ``seed model" as input and iteratively ``grooms" the disk until the 3D dust distribution matches that of a steady-state collisional disk with a given dust production rate.  A more detailed explanation of the collisional grooming algorithm can be found in \citet{sk09}.

To produce the collisionless seed models, we simulated the dynamical interactions between a cloud of dust and a single planet on a circular orbit around a Sun-like star.  We investigated four different planet masses, $M_p = [1 $ M$_{\rm E}$, 5 M$_{\rm E}$, 1 M$_{\rm N}$, 1 M$_{\rm J}$], where M$_{\rm E}$, M$_{\rm N}$, and M$_{\rm J}$ are the mass of the Earth, Neptune, and Jupiter, respectively.  Each of these planet masses were modeled with two values for orbital semi-major axis, $a_p = [0.5, 1.0]$ AU.  For the Jupiter-mass case we also created models with $a_p = 0.1$ AU.

For each of these models, we integrated the orbits of 12,500 dust grains launched from a belt of parent bodies exterior to the orbit of the planet.  We used a hybrid symplectic integrator with an integration step size equal to one hundredth of the period of the planet \citep{sk08}.  The parent body orbits were distributed uniformly in semi-major axis between $2.5 a_p$ and $3.0 a_p$, uniformly in eccentricity between 0 and 0.1, and uniformly in inclination between 0 and 10$^{\circ}$.  All other parent body orbital parameters were distributed uniformly between 0 and $2\pi$.

For the seed model integration, we distributed the dust grains in equal numbers among 25 logarithmically-spaced $\beta$ values, ranging from 0.43355 to 0.00046, where $\beta$ is the ratio of the radiation pressure force on a grain to the stellar gravitational force.  Each discrete $\beta$ value represents the behavior of a small range of $\beta$ values from $\beta/\sqrt{1.33}$ to $\beta\sqrt{1.33}$, so that we represent grains down to the blowout size ($\beta = 0.5$).  These $\beta$ values correspond to 25 values of grain size, ranging roughly from sub-micron to nearly millimeter sizes.  We calculated $s(\beta)$ according to
\begin{equation}
	s = 0.57\, \frac{\langle Q_{\rm PR} \rangle}{\beta}\, \frac{1 {\rm\; g\; cm}^{-3}}{\rho}\;\; {\micron},
\end{equation}
valid for a Sun-like star, where $\rho$ is the mass density of a dust grain and $\langle Q_{\rm PR} \rangle$ is the Poynting-Robertson efficiency, $Q_{\rm PR}$, averaged over the solar spectrum \citep{bls79}.  We let $\rho = 2$ g cm$^{-3}$, appropriate for ``rocky" grains.  $Q_{\rm PR}$, an implicit function of $s$, is given by
\begin{equation}
	Q_{\rm PR} = Q_{\rm abs} + Q_{\rm sca} \left( 1 - \langle \rm{cos}\; \alpha\rangle\right),
\end{equation}
where $Q_{\rm abs}$ and $Q_{\rm sca}$ are the absorption and scattering efficiencies of the dust grain and $\langle \rm{cos}\; \alpha\rangle$ is the cosine of the scattering angle weighted by the scattering phase function and averaged over all directions \citep{bls79}.

Values for $Q_{\rm abs}$ and $Q_{\rm sca}$, as well as the scattering phase function for exozodiacal dust grains are unknown.  Observations of dust in the outer regions of debris disks reveal grains that appear to deviate significantly from Mie theory \citep[e.g.][]{kgc05,k10}.  In light of this, we chose simple generic laws for these quantities and investigated the effects of varying the dust grains' scattering parameters.  We let $Q_{\rm abs} = 1$ for $\lambda \le 2 \pi s$ and $Q_{\rm abs} = (2 \pi s / \lambda)^2$ for  $\lambda > 2 \pi s$.  We used three values for the albedo, $\omega = [0.1, 0.3, 0.5]$, where $\omega = Q_{\rm sca} / (Q_{\rm sca} + Q_{\rm abs})$.  Finally, we used a Henyey-Greenstein scattering phase function \citep{hg41}, given by
\begin{equation}
	P\left(\alpha\right) = \frac{1}{4\pi} \frac{1 - g^2}{\left( 1 + g^2 - 2g \cos{\alpha} \right)^{3/2}}\; ,
\end{equation}
and investigated three values for the scattering asymmetry parameter $g = [0.2, 0.5, 0.8]$.  Since $g = \langle \cos{\alpha}\rangle$, $g=0$ corresponds to isotropic scattering and increasing values of $g$ in the range $0<g<1$ correspond to increasing degrees of forward scattering.

We recorded the positions and velocities of the grains in the frame co-rotating with the planet every $t_{\rm PR} / 14300$ years for a maximum time of 1 Gyr.  This time between records resolves the collision time for the smallest grains in our 500 zodi simulations by a factor of $\sim10^4$.  During the integration, grains were removed if their semi-major axis $a<a_p/3$ or $a>300$ AU, if they collided with the planet given realistic planet radii, or if they were scattered into hyperbolic orbits and ejected from the system.  For the case of $M_p = 5$ M$_{\rm E}$, we set the planet radius equal to 1.5 Earth radii, appropriate for a roughly Earth-like composition \citep{skh07}.  For all other planets we used the radii of their respective solar system counterparts.

We then processed the collisionless seed models with the collisional grooming algorithm to obtain steady-state collisional disk models.  We weighted the dust production rate of each grain size such that at the moment of launch from a planetesimal the dust grains were produced according to a Dohnanyi crushing law, $dN/ds \propto s^{-3.5}$ \citep{d69}.  The collisional grooming algorithm then solved the number flux equation and self-consistently determined the size distribution at all points in the disk.

For each collisionless seed model, we produced 6 different collisional disk models with increasing optical depths (i.e. increasing disk masses/dust production rates).  We chose the total dust production rates such that the 6 collisional disk models had \emph{maximum} optical depths of $10^{-7}$, $5 \times 10^{-7}$, $10^{-6}$, $2 \times 10^{-6}$, $5 \times 10^{-6}$, and $10^{-5}$.  For the Jupiter-mass case we also modeled a maximum optical depth of $5 \times 10^{-5}$.  These maximum optical depths roughly correspond to dust production rates of $2.6 \times 10^{-10}$, $2.3 \times 10^{-9}$, $7.1 \times 10^{-9}$, $2.1 \times 10^{-8}$, $8.5 \times 10^{-8}$, $2.2 \times 10^{-7}$, and $2.0 \times 10^{-6}$ M$_{\rm E}$ Myr$^{-1}$ for dust grain sizes ranging from the blowout size up to $\sim$1 mm in radius, though dynamical differences can cause the dust production rate to vary by as much as a factor of 2 among models with the same optical depth.  Here ``optical depth" refers to the dust grain cross section per unit area of the disk integrated along a line normal to the disk plane and does not directly depend on the dust grains' optical constants.  To mitigate the effects of Poisson noise when determining the maximum optical depth, we azimuthally averaged the disks' optical depths into radial bins of width $0.05 a_{\rm p}$.  For the rest of this paper, we will use 1 ``zodi" to refer to a maximum optical depth of $10^{-7}$; our models range from 1 zodi to 500 zodis of dust.

We made sure that the collisional grooming algorithm resolved any fine structure within the disk by adopting a 512 $\times$ 512 $\times$ 128 grid with a bin size of $0.05a_p$ $\times$ $0.05a_p$ $\times$ $0.025a_p$.  We ran the collisional grooming algorithm until no single dust grain record changed by more than 5$\%$ from one iteration to the next.  \citet{sk09} performed a number of tests showing that this criterion is sufficient to achieve convergence to the solution of the number flux equation for disks with and without azimuthal asymmetries.  This degree of convergence is roughly equivalent to an uncertainty of less than 0.05$\%$ in the bin-averaged density distribution for the densest regions of our models.

Similar to the collisional models of the Kuiper Belt dust disk in \citet{ks10}, our models draw a distinction between cratering events and catastrophic collisions.  Catastrophic collisions occur when the collisional energy measured in the center of momentum frame of the two dust grains exceeds the critical disruption energy of the dust grains.  In this case, both grains are destroyed.  All other events may be deemed cratering events, which we ignore.  In these cases, both grains remain unchanged.  In the strength regime the critical disruption energy is commonly approximated by
\begin{equation}
	Q = A_s \left( \frac{s}{\rm{1\; m}} \right)^{b_s}.
\end{equation}
We used the parameters for ``rocky" grains from \citet{kls06} and let $b_s=-0.24$ and $A_s=10^6$ erg g$^{-1}$.

We note that our treatment of collisions ignores fragments.  Although this may be a decent approximation for icy grains produced in the outer regions of the solar system as discussed in \citet{ks10}, the validity of this approximation is not as clear for the grains simulated here.  We discuss this approximation further in Section \ref{caveats_section}.

\subsection{Synthesizing Exozodi Light Curves}

For each of the above models, we used our publicly-available IDL software package \emph{dustmap} to synthesize edge-on images of the dust disk as a function of planetary phase.  Given a user-specified disk orientation and distance, \emph{dustmap} illuminates the disk with starlight and calculates the scattered and emitted light from each individual dust grain in the disk model.  The code is capable of calculating the scattering and absorption efficiencies of the dust as a function of composition, grain size, and wavelength using Mie theory, but we chose to use the generic values described in Section \ref{exozodimodels}.  For thermal emission images, \emph{dustmap} uses the stellar spectrum, approximated as a blackbody, and the wavelength-dependent dust absorption efficiency to calculate the equilibrium temperature of the dust as a function of grain size and circumstellar distance.  For scattered light images, \emph{dustmap} determines the scattering angle for each dust grain and uses a Henyey-Greenstein phase function to calculate the flux scattered toward the observer.

When calculating the contribution of the stellar flux, our imaging code takes into account extinction of the stellar light by transiting dust, but does not include the effects of stellar limb darkening.  For transits of optically-thick objects with well-defined edges, such as planets, these effects are important.  However, the clumps in an optically thin exozodiacal cloud have comparatively smooth features; the projected angular size of the dust clump's ``edge" is typically larger than the angular size of the star, rendering limb darkening effects unimportant.

We imaged our model disks at a wavelength of 575 nm, the peak in \emph{Kepler}'s response curve \citep{kbb10}.  We use our images at this single wavelength to approximate the exozodi signal integrated across the entire \emph{Kepler} photometric response curve, which extends roughly from 425 nm to 900 nm.  This approximation is valid given our approximations for $Q_{\rm abs}$ and $Q_{\rm sca}$.

We produced transit light curves from our models by calculating the total flux at 72 values of planetary phase equally spaced over 360$^{\circ}$.  The relative intensity of our transit light curves was calculated as
\begin{equation}
	I \left( \phi \right) = \frac{F \left( \phi \right)}{F_{\rm max}},
\end{equation}
where $F(\phi)$ is the total (disk plus stellar) flux as a function of planetary phase $\phi$, $F_{\rm max}$ is the maximum of $F(\phi)$, and the stellar flux was calculated assuming a 1 solar radius spherical blackbody with an effective temperature of 5778 K.

\section{Results\label{results_section}}

\subsection{Resonant rings in collisional disks}

\begin{figure}
\begin{center}
\includegraphics[height=6.5in]{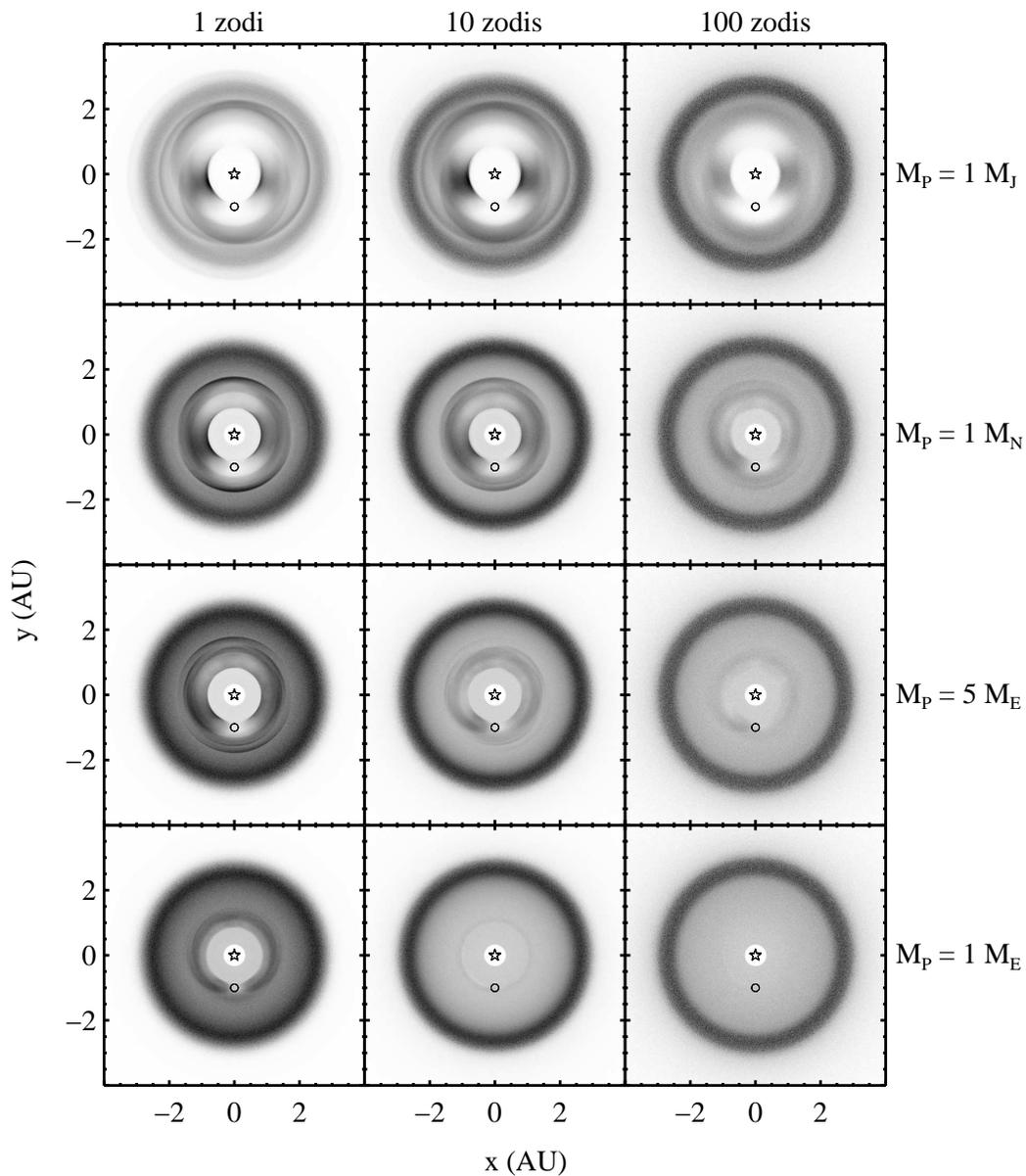}
\caption{Face-on optical depths for twelve of our models.  Each optical depth histogram is scaled independently to show detail.  For all models shown, $\omega = 0.3$, $g=0.5$, and $a_p=1.0$ AU.  Each row corresponds to one of the four modeled planet masses, as labeled along the right side.  Each column corresponds to one of the maximum optical depths modeled, as labeled along the top of the figure.  At optical depths $\sim 100$ zodis, a Jupiter-mass planet can create significant resonant asymmetry while an Earth-mass planet cannot.\label{disk_od_figure}}
\end{center}
\end{figure}

Figure \ref{disk_od_figure} shows face-on optical depth histograms for twelve of our models.  Each row corresponds to one of the four modeled planet masses and each column corresponds to one of the seven modeled zodi levels.  Each model is scaled independently to show detail.  The location of the star and planet are marked with a star and a circle, respectively, and the planets orbit counterclockwise.  The birth ring, which traces the location of the planetesimals, appears as a symmetric circular ring between 2.5 and 3.0 AU.  Resonant ring structures appear interior to the birth ring, near 1 AU.

The 1 zodi models (left column) illustrate the features typically attributed to resonant ring structures: a gap near the location of the planet, an overdense ``clump" trailing the planet, and in some cases a slightly less overdense clump leading the planet in its orbit.  The resonant ring created by a Jupiter-mass planet is clearly more dense and asymmetric than the resonant ring created by an Earth-mass planet.  This is because the Jupiter-mass planet traps dust more efficiently, thereby trapping more dust and trapping a larger fraction of that dust in the 2\,:\,1 resonance, which contributes strongly to the structure's bilobed appearance.

The rows in Figure \ref{disk_od_figure} illustrate the effects of collisions on resonant ring structures.  As previously described in \citet{sk09} and \citet{ks10}, increasing the optical depth of a disk increases the collision rate, which in turn erases resonant ring structure.  However, the top-right histogram in Figure \ref{disk_od_figure} shows that even for an optical depth of 100 zodis, a Jupiter-mass planet can create significant resonant asymmetry in a disk.

\subsection{Light curves\label{lightcurves_section}}

\begin{figure}
\begin{center}
\includegraphics[height=4.0in]{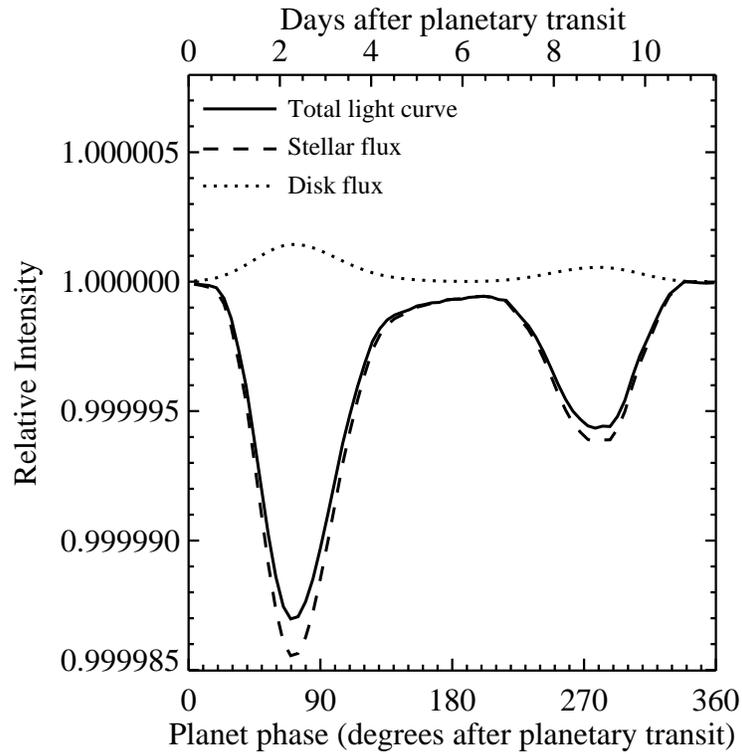}
\caption{Light curve for a 10 zodi dust disk with $\omega=0.5$ and $g=0.8$, perturbed by a Jupiter-mass planet on a circular orbit with $a_p=0.1$ AU.  Transiting dust clumps create the broad minima.  The total light curve is the result of a competition between occulted stellar flux and forward-scattered flux from the disk.  \label{lightcurve_figure1}}
\end{center}
\end{figure}

Figure \ref{lightcurve_figure1} shows a sample light curve for a 10 zodi disk perturbed by a Jupiter-mass planet.  The planetary transit occurs at a planet phase of $0^{\circ}$, but is not shown in this plot because it is 3 orders of magnitude larger than the disk transit and lasts for only a few hours.  The trailing and leading dust clumps in the disk transit at approximately $90^{\circ}$ and $270^{\circ}$, respectively.  The solid line in Figure \ref{lightcurve_figure1} shows the total light curve for the star/disk system, the result of a competition between stellar extinction and forward scattering from the disk.

The dashed line shows the variation due to stellar extinction alone.  Similar to a planetary transit, the enhanced column density of dust between the observer and the star during the transit of a dust clump reduces the amount of starlight that reaches the observer.  However, unlike a planetary transit, the projected angular size of the light-blocking material (the clump) and the angular scale of the clump's ``edges" are typically much larger than the angular size of the star.  This difference in angular size produces transit minima that are much broader than that of a planet. 

The dotted line in Figure \ref{lightcurve_figure1} show the variations in the disk flux alone.  Since the angular size of a clump is much larger than the angular size of the star, dust just pre- or post-transit forward-scatters additional starlight toward the observer.  This effect produces a disk flux that peaks when the stellar extinction reaches a minimum.  

The forward scattering effect for our $g=0.8$ model, shown by the dotted line in Figure \ref{lightcurve_figure1}, may be unrealistically large.  The outer regions of observed debris disks exhibit $g$ values typically much smaller, on the order of 0.2 or less \citep{alm99, kgc05, dws08, k10}.  \citet{h85} fit the scattering phase function of the zodiacal cloud using a weighted series of three Henyey-Greenstein phase functions and found the dominant term to be highly forward scattering, with $g=0.7$.  However, this assumes a dust distribution $\propto r^{-1}$, and when correcting for the $r^{-1.34}$ distribution determined by \citet{k98}, the effective $g$ value decreases significantly.

\begin{figure}
\begin{center}
\includegraphics[height=4.0in]{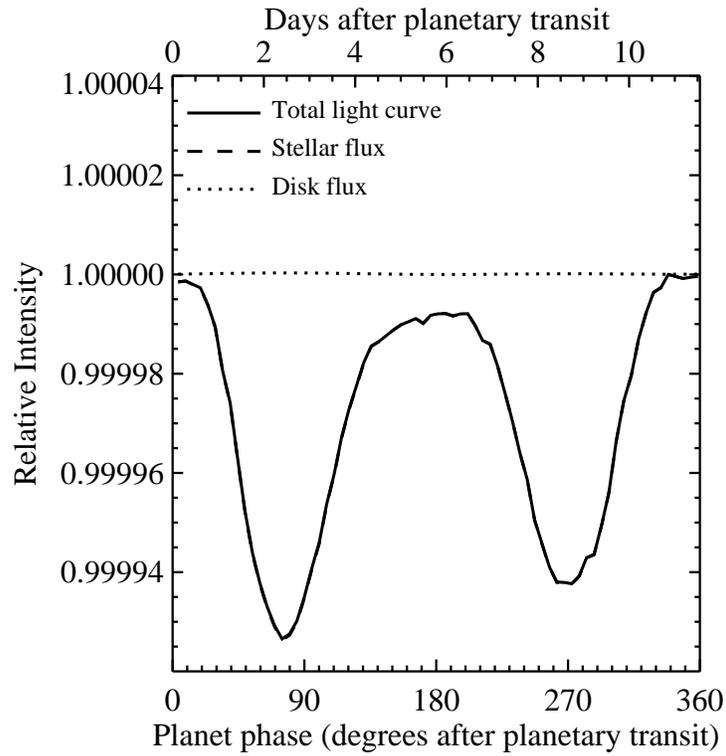}
\caption{Light curve for a 100 zodi disk with $g = 0.2$ and all other parameters the same as in Figure \ref{lightcurve_figure1}. The transit depth amplitude is approximately $7 \times 10^{-5}$.  The trailing and leading clumps have similar optical depths, a result of small grains being preferentially trapped to librate about the leading center of the 2\,:\,1 resonance. \label{lightcurve_figure2}}
\end{center}
\end{figure}

We produced other models with lesser degrees of forward scattering ($g=0.2$, 0.5).  Forward-scattered light from the disk contributes negligibly to the light curves for all of these models.  Figure \ref{lightcurve_figure2} shows an example light curve for such a model.  This model is the same as that shown in Figure \ref{lightcurve_figure1}, except $g=0.2$ and the disk density was increased to 100 zodis.

\subsection{Asymmetry reversal and light curve complexities\label{asymmetry_section}}

The light curves shown in Figure \ref{lightcurve_figure1} and Figure \ref{lightcurve_figure2} also highlight the morphological effects of collisions on resonant ring structures.   The 10 zodi model shown in Figure \ref{lightcurve_figure1} shows a trailing clump that is significantly more dense than the leading clump, while the 100 zodi model shown in Figure \ref{lightcurve_figure2} shows a trailing clump density on par with the leading clump density.  At $a_p=1.0$ AU, the changes are even more significant, as shown in the upper row of Figure \ref{disk_od_figure}.  In this case, the leading clump is slightly more dense than the trailing clump for a disk density equal to 100 zodis.

\begin{figure}
\begin{center}
\includegraphics[height=3.0in]{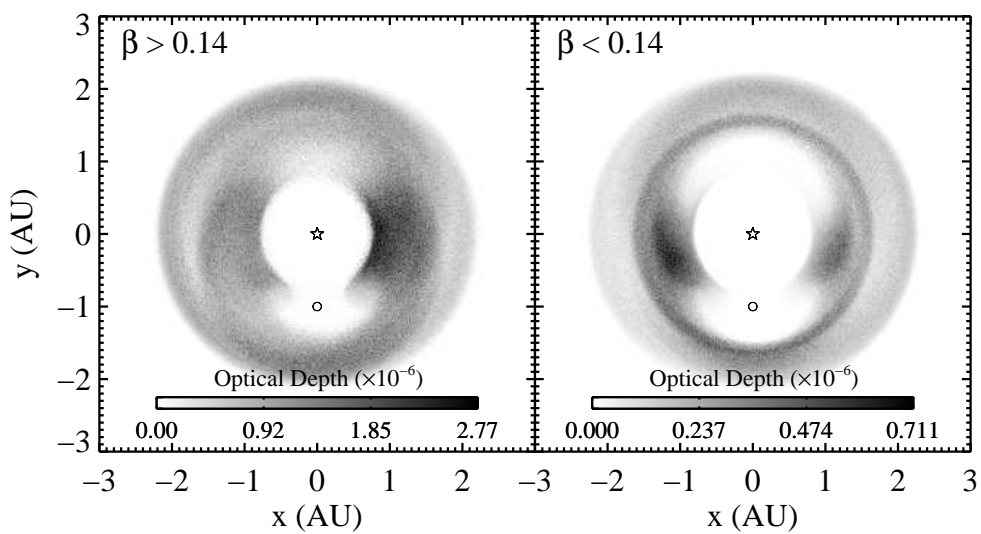}
\caption{Face-on optical depth of all grains with semi-major axes near the semi-major axis region of the 2\,:\,1 MMR for a 100 zodi simulation with $M_p = 1$ M$_{\rm J}$ and $a_p = 1.0$ AU.  The planet, marked with a circle, orbits counterclockwise.  The smallest grains ($\beta > 0.14$) create a clump that leads the planet in its orbit while the largest grains ($\beta < 0.14$) create a clump that trails the planet.  In this case, the small grains dominate the optical depth of the disk, leading to asymmetry reversal with a denser leading clump. \label{asymmetry_figure}}
\end{center}
\end{figure}

It appears that collisions are ``choosing" to remove grains from the trailing clump, but a more detailed look at the optical depth as a function of grain size reveals that this effect stems from resonant dust grain size-sorting.  Figure \ref{asymmetry_figure} shows the optical depth for all dust grains (resonantly trapped and non-trapped) within the region of the 2\,:\,1 MMR for the case of a Jupiter-mass planet orbiting at 1.0 AU in a 100 zodi dust disk.  These dust grains were selected by their semi-major axes $a$ such that $0.95\, a_{2:1} < a < 1.05\, a_{2:1}$ where $a_{2:1} = a_p\,2^{2/3}(1-\beta)^{1/3}$.  The largest grains ($\beta < 0.14$) create a dust clump that trails the planet while the smallest grains ($\beta > 0.14$) create a dust clump that leads the planet.  Figure \ref{asymmetry_figure} shows that for grains within the region of the 2\,:\,1 MMR, the leading dust clump contributes roughly 50\% more to the optical depth than the trailing dust clump.  Including grains at all semi-major axes reduces this asymmetry reversal to approximately 5\%.  A close examination of the paths of individual particles reveals that this asymmetry reversal does not suffer from small number statistics; roughly 350 dust grains with $\beta>0.14$ occupy orbits that contribute to a denser leading clump while 270 contribute to a denser trailing clump.

This size-dependent asymmetry extends to disks with much lower optical depths, $\sim 1$ zodi.  To first order, collisions simply select which grain size dominates the optical depth of the resonant ring structure.  For lower disk optical depths $\sim 1$ zodi, the PR times of grains with $\beta < 0.14$ can be less than their collision times.  In this case, large grains migrate inward and reach the planet's MMRs to create a trailing clump that dominates the resonant ring's optical depth.  For denser disks, like that shown in Figure \ref{asymmetry_figure}, the grain-grain collision time is reduced and only grains with $\beta > 0.14$ can migrate inward to populate the planet's MMRs, creating a dominant leading clump.

The leading-trailing clump asymmetry only reverses for the $M_p=1$ M$_{\rm J}$ models.  These models are the only models for which $j$\,:\,1 exterior resonances dominate the MMR distribution, and the $j$\,:\,1 resonances are unique in that they can lead to asymmetric libration.  Asymmetric libration arises from a balance between the direct and indirect terms of the disturbing function and splits the libration center of a resonance into two separate libration centers for particles that obtain moderate orbital eccentricities.  These libration centers lead and trail the planet by $\sim90^{\circ}$ for the $M_p=1$ M$_{\rm J}$ case.  It is useful to refer to a single $j$\,:\,1 MMR with asymmetric libration as two separate resonances, which we will refer to as leading and trailing resonances.

The leading and trailing $j$\,:\,1 resonances have different trapping efficiencies.  Several authors have found that in the case of a planet migrating outward toward a collection of massless particles, particles trapped in the 2\,:\,1 resonance typically populate the leading and trailing resonances equally or preferentially populate the trailing resonance, creating a dominant clump of particles trailing the planet \citep{w03, mcc05}.  However, in their simulations of the outward migration of Neptune in the Solar System, \citet{mcc05} showed that for some migration rates the particles preferentially populated the leading resonance by up to a factor of 2 (see their Figures 15 and 16), creating an asymmetry reversal similar to what we observe in our models.  \citet{mcc05} attributed this asymmetry reversal to an asymmetric shift in the stable and unstable equilibrium points for the resonant angle such that they form a relatively flat potential near the leading center.  As a result, the particles spend more time traversing the region near the leading center where the libration rate is much lower, leading to a higher probability of capture into the leading resonance \citep{mcc05}.  Although our models feature particles spiraling in toward a non-migrating planet, the physics of asymmetric libration is similar to the case of outward planetary migration \citep{q06, mw11} and the origin of the asymmetry is likely identical to that described by \citet{mcc05}.

The asymmetry reversal seen in the $j$\,:\,1 resonances demonstrates the size-sorting effects of resonances.  For example, for the case of a Jupiter-mass planet, we expect to find smaller grains in the leading clump and larger grains in the trailing clump.  These clumps may therefore exhibit differing spectral features, and differences in their color may possibly be detectable with future multi-band resolved imaging or multi-band transit photometry.

\begin{figure}
\begin{center}
\includegraphics[height=4.0in]{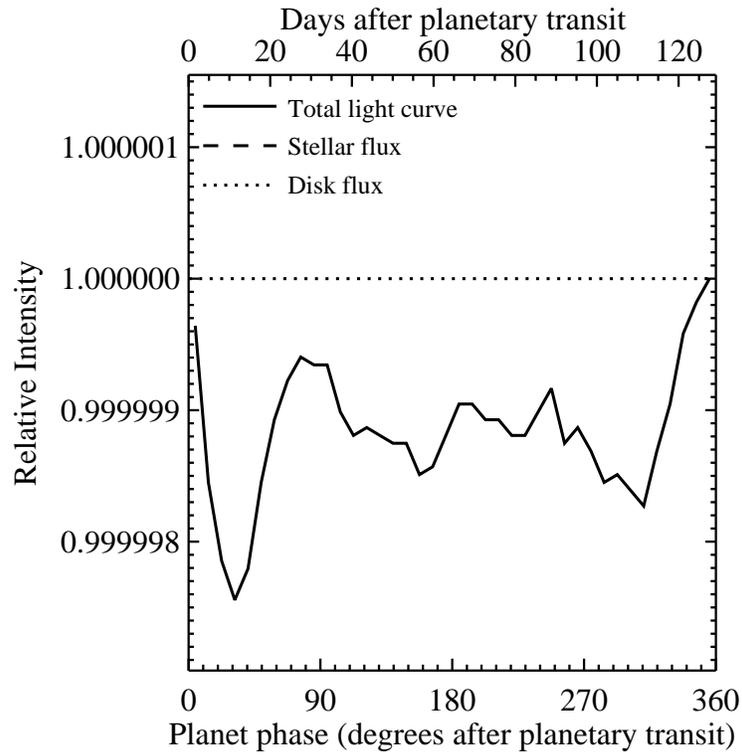}
\caption{Light curve for a 10 zodi disk with $\omega = 0.5$ and $g=0.2$ perturbed by a 5.0 M$_{\rm E}$ planet on a circular orbit at 0.5 AU.  The minimum near $180^{\circ}$ is real, created by a density enhancement in the resonant ring.  Variations of amplitude $\sim10^{-7}$ can be attributed to Poisson noise.\label{lightcurve_figure3}}
\end{center}
\end{figure}

The transit light curves of resonantly trapped dust are not always composed of two distinct minima.  Figure \ref{lightcurve_figure3} shows the light curve for a disk with identical parameters to that shown in Figure \ref{lightcurve_figure2}, but perturbed by a 5.0 M$_E$ planet on a 0.5 AU orbit.  The $\sim10^{-7}$ variations in the light curve are attributable to Poisson noise from the model, but larger variations are real.  In this case, the resonant ring structure does not resemble a simple bilobed structure (see Figure \ref{disk_od_figure}); the multiple, subtle clumps in the resonant ring produce several smaller minima.  

\subsection{Light curve amplitudes\label{amplitudes_section}}

We would like to know the maximum signal amplitude that exozodiacal resonant ring structures can create.  To determine this, we calculated the amplitude $A$ of each modeled light curve, given simply by $A = (1.0 - I_{\rm min})$.

As alluded to by Figure \ref{lightcurve_figure3}, some modeled light curves suffer significantly from Poisson noise.  To avoid amplitudes that were dominated by Poisson noise, we fit each light curve with an eighth-order polynomial and then calculated the residual standard deviation.  We rejected curves with amplitudes less than four times the residual standard deviation.  We re-synthesized the rejected curves using a finer sampling for the planet phase (2000 values for planet phase as opposed to the original 72) and then binned the light curves down to 50 planet phase values to mitigate Poisson noise.  We fit these curves again with an eighth order polynomial, and permanently rejected those with amplitudes less than four times the residual standard deviation.  In all cases, the final rejected light curves failed our Poisson noise test because relatively little resonant ring structure existed within the disk.

\begin{figure}
\begin{center}
\includegraphics[height=4.0in]{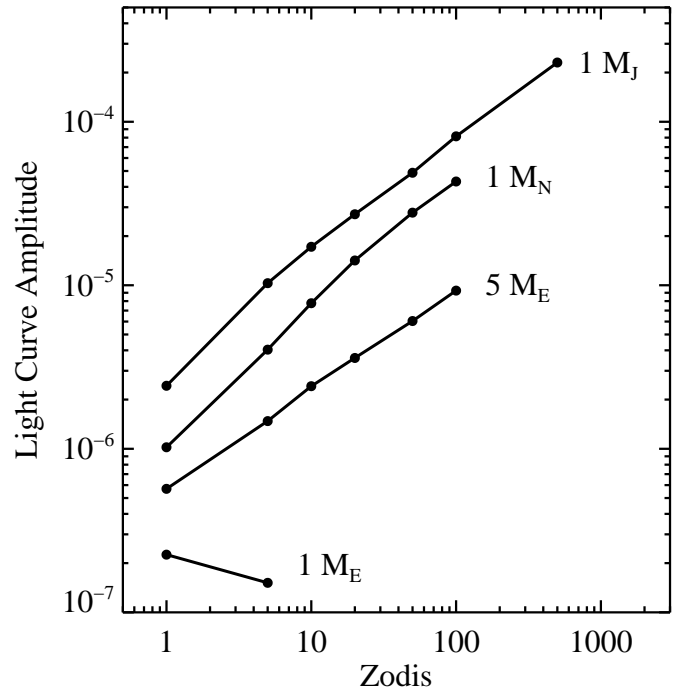}
\caption{Light curve amplitude as a function of zodi level for all four modeled planet masses.  In all cases $a_p$ = 0.5 AU, $\omega = 0.5$, and $g = 0.2$.  Jupiter-mass planets can create resonant disk structures with maximum transit depths $\sim 10^{-4}$ while Earth-mass planets cannot create detectable resonant disk structures. \label{amp_vs_zodis_figure}}
\end{center}
\end{figure}

Figure \ref{amp_vs_zodis_figure} shows the results for the $a_p=0.5$ AU case with $\omega=0.5$ and $g=0.2$, although the results are nearly independent of $g$.  Only two data points are shown for the $M_p = 1$ M$_E$ case.  The amplitudes at larger disk densities did not meet our Poisson noise criteria described above; an Earth-mass planet creates a very weak resonant ring structure in disks with optical depths greater than a few tens of zodis.

As expected, the light curve amplitude increases with disk density.  However, the amplitude of the signal does not scale linearly with the number of zodis; for the 1 M$_J$ curve, the maximum signal amplitude for the 100 zodi case is only $\sim$40 times greater than the 1 zodi case.  We ascribe this non-linearity to two effects brought about by grain-grain collisions.  First, the enhanced collision rate in denser disks will reduce the fraction of dust that migrates inward from the planetesimal belt to populate the planet's MMRs \citep{w06, sk09}.  Second, \citet{sk09} showed that grain-grain collisions preferentially remove dust from MMRs and smooth out disk asymmetries.  These phenomena effectively curb the signal amplitude as the disk density increases.

The only exception to this rule is the light curve amplitude for a resonant ring created by an Earth-mass planet, which decreases with increasing optical depth.  In general, trapping efficiency decreases as the dust grain size decreases \citep[e.g.][]{w03}, and increasing the optical depth reduces the typical dust grain size that reaches the planet's resonances \citep{w06, sk09}.  In the Earth-mass case, the planet traps dust so inefficiently that even a moderate optical depth results in grains that are typically too small for the planet to trap.  The bottom row in Figure \ref{disk_od_figure} shows clearly that an Earth-mass planet traps relatively little dust into resonance for optical depths greater than a few zodis.

Figure \ref{amp_vs_zodis_figure} shows that disks perturbed by planets less massive than Neptune are unlikely to produce a signal detectable with \emph{Kepler}.  Disks perturbed by Neptune-mass planets can produce signals greater than $10^{-5}$ in amplitude, while dust trapped by super-Earths can only exceed the $10^{-6}$ level.  Resonant rings created by Earth-mass planets have little hope of being detected via transit photometry, with signal amplitudes on the order of $10^{-7}$ or less.

Figure \ref{amp_vs_zodis_figure} also shows that exozodiacal dust resonantly trapped by a single Jupiter-mass planet on a circular orbit can produce a transit signal that exceeds the $10^{-4}$ level for disks with optical depths of several hundred zodis.  To within a factor of a few, this was the largest amplitude we could produce with our models.  We found that even larger zodi values could increase the scattered disk flux signal by a factor of a few, but disks with optical depths greater than $\sim1000$ zodis at a few tenths of an AU have shorter lifetimes and are therefore less likely to be detected (see discussion of disk lifetimes in Section \ref{disklifetimes_section}).

We briefly examined the effects of increasing the planet mass beyond 1 Jupiter mass to $M_p=2$ M$_{\rm J}$ and 5 M$_{\rm J}$, and found that the signal amplitude \emph{decreased}.  By examining the semi-major axis distribution of dust grains, we determined two causes for this decrease.  First, increasing the planet mass beyond a Jupiter mass improves the trapping efficiency of higher-order MMRs (e.g. 3\,:\,1 and 5\,:\,2).  Since resonant trapping pumps up the dust grains' orbital eccentricities, the increased population of grains in higher-order MMRs serves to enhance the collision rate at larger circumstellar distances.  This in turn reduces the number of grains available to migrate inward to the 2\,:\,1 resonance, where trapping is most efficient.

\begin{figure}
\begin{center}
\includegraphics[height=4.0in]{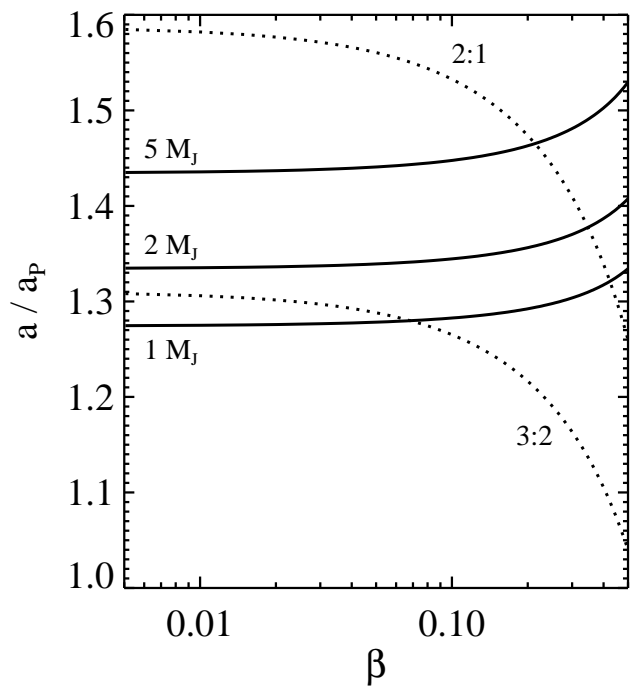}
\caption{Chaotic zone boundaries (solid lines) as functions of $\beta$ for $M_p = 1$, 2, and 5 M$_{\rm J}$, plotted with the locations of the 2\,:\,1 and 3:2 MMRs (dotted lines). Sustained resonant trapping can occur when the MMR semi-major axis is greater than the chaotic zone boundary.  As planet mass increases the 2\,:\,1 resonance intersects the chaotic zone boundary at smaller values of $\beta$; more massive planets trap fewer small grains into the 2\,:\,1 resonance.  \label{chaos_vs_resonance_figure}}
\end{center}
\end{figure}

Second, increasing the planet mass moves the exterior chaotic zone boundary to larger semi-major axes.  The exterior chaotic zone boundary, expressed as
\begin{equation}
	a_{\rm chaotic} = a_p \left( 1 + \kappa \left( \frac{\mu}{1- \beta} \right)^{2/7} \right),
\end{equation}
is a semi-major axis boundary interior to which MMRs overlap and sustained resonant trapping is unlikely \citep{w80}.  Estimates for the value of $\kappa$ vary; we chose the most recent estimate, $\kappa = 2.0$, which comes from numerical simulations of the Fomalhaut debris ring by \citet{ckk09}.  If the semi-major axis of an exterior $j$\,:\,$k$ MMR, given by
\begin{equation}
	a_{j{\rm :}k} = a_p \left(\frac{j}{k}\right)^{2/3} \left(1-\beta\right)^{1/3},
\end{equation}
is less than $a_{\rm chaotic}$, then the resonance will not be significantly populated.  Figure \ref{chaos_vs_resonance_figure} shows the semi-major axes for the 2\,:\,1 and 3:2 MMRs as functions of $\beta$, along with the chaotic zone boundaries for several planet masses.  As Figure \ref{chaos_vs_resonance_figure} shows, for $M_p=1$ M$_{\rm J}$ the 2\,:\,1 MMR is exterior to the chaotic zone boundary for $\beta < 0.4$, but for $M_p = 5$ M$_{\rm J}$ only grains with $\beta \lesssim 0.2$ can be trapped in the 2\,:\,1 MMR, significantly reducing the population of the 2\,:\,1 MMR in the latter simulation.

We also examined the amplitudes of the variable disk flux signal, shown as dotted lines in Figures \ref{lightcurve_figure1}, \ref{lightcurve_figure2}, and \ref{lightcurve_figure3}.  Disks inclined from an edge-on orientation by an amount greater than their disk opening angle would not exhibit the broad transit minima shown in these figures, but would still feature the variable disk flux caused by forward-scattering.  Our models predict that the largest possible forward-scattered light signal that a disk with a ring of resonantly-trapped dust can produce is roughly $5\times10^{-6}$.  Detecting resonantly-trapped dust rings in disks that are not in an edge-on orientation would therefore require photometric precision better than one part in $10^{6}$ over the course of an orbital period.  Additionally, to produce such a large scattered light signal the disk must have an optical depth of at least 100 zodis, strongly forward scattering dust ($g=0.8$) with a large albedo ($\omega=0.5$), a resonant ring structure created by a Jupiter-mass planet just exterior to the sublimation distance for silicate grains, and an inclination just greater than the disk opening angle.  Simultaneously satisfying all of these criteria may only occur very rarely and we consider this scenario unlikely.

\section{Discussion\label{discussion_section}}

\subsection{Disk lifetimes\label{disklifetimes_section}}

Figure \ref{amp_vs_zodis_figure} shows that the amplitude of a resonant dust ring's light curve approaches the $10^{-4}$ level only for massive perturbing planets, $\sim 1$ M$_{\rm J}$, with disk optical depths $\sim 100$ zodis.  Additionally, the likelihood of photometrically detecting transiting dust clumps improves as the planet semi-major axis decreases, since the transit probability increases, the number of orbit-foldings of the light curve increases for a given total observation time, and the required timescale for photometric stability decreases.  Radial velocity and transit measurements have already shown a wealth of close-in massive planets, but should we also expect to find hundreds of zodis of dust?

Massive amounts of hot dust ($> 1000$ zodis) have already been observed around other stars \citep[e.g.][]{b05, szw05, wgd05, adm06, rsz08, acm09, aml09, map09, mzr10, wbs11, mgsm11}, although the frequency of these disks around mature ($>1$ Gyr) solar-type stars is only on the order of a few percent \citep{b06}.  Observations have also revealed a few less massive exozodiacal disks with optical depths greater than a few hundred zodis \citep{map09, mgsm11}.  The least massive of these detected exozodiacal clouds are consistent with the long term collisional erosion of a belt of planetesimals \citep{wsg07, map09}, similar to the disks we modeled.  But the dust in many of these systems is likely transient in nature; current estimates restrict the typical disk mass $e$-folding times to less than $\sim 10$ Myr, much less than the age of many of these systems \citep{wsg07}.  

However, transient collisions may also produce transient resonant rings.  After all, the fundamental mechanism behind our models is very basic: collisions produce dust, some fraction of which migrates inward and becomes trapped in a planet's mean motion resonances.  We suspect that a massive collisional avalanche initiated by a single event ends with the inward migration of any remaining bound grains.  Future modeling of massive, transient collisions is necessary to understand the likelihood of this scenario.

The lifetime of dense exozodiacal disks is not well known, but \citet{wsg07} place loose upper limits on the age of these disks.  Using Equation 21 in \citet{wsg07}, along with the fact that the disk lifetime could be two orders of magnitude longer (see Section 3.2 in \citet{wsg07}), we calculate maximum disk lifetimes of 1-100 Myr, 34-3400 Myr, and 0.17-17 Gyr for our $a_p=0.1$, 0.5, and 1.0 AU 100 zodi disk models, respectively.  These calculated lifetimes vary with $a_p$ and suggest that disks with 100 zodis of dust at distances less than $\sim$1 AU only exist around the youngest stars, while disks with 100 zodis of dust at distances larger than $\sim$1 AU may exist for up to billions of years.  However, the lifetime varies with $a_p$ only because our models assumed planetesimals distributed from $2.5a_p$ to $3.0a_p$; for the case of $a_p=0.1$ AU the assumed planetesimals are on short-period orbits near 0.3 AU.  In reality, dust may drift inward to 0.1 AU from a belt of planetesimals located farther out where the disk is significantly longer-lived.

The model shown in the lower right panel of Figure \ref{disk_od_figure} clearly illustrates how a more distant source of planetesimals may deliver copious amounts of dust to the very inner regions of a planetary system.  In this model planetesimals near 3 AU produce 100 zodis of dust.  The dust migrates inward past the Earth-mass planet at 1 AU, which has little effect on the dust distribution.  The optical depth interior to 1 AU (but exterior to our integration-imposed inner cutoff of 0.3 AU) is approximately constant as a function of circumstellar distance; nearly all of the dust that migrates to 1 AU would continue to migrate to 0.1 AU.

Dense exozodiacal clouds may be fed by even more distant sources of dust, e.g. massive Kuiper Belt analogs, for which the disk's lifetime would be of no concern.  Both \citet{ks10} and \citet{vkl10} simulated the distribution of dust produced by the Kuiper Belt using models that included grain-grain collisions.  \citet{ks10} and \citet{vkl10} both showed that small Kuiper Belt dust grains can migrate into the inner solar system via PR drag, even if the Kuiper Belt dust disk is dense enough to be in the ``collision-dominated" regime.  \citet{ks10} estimated that $\sim60$ zodis of dust would make it to 1 AU from a 1000-zodi Kuiper Belt analog in the absence of Saturn and Jupiter.  \citet{sc06} showed that strong stellar winds can also transport large amounts of dust into the inner disk in spite of grain-grain collisions; an AU Mic disk-analog could deliver more than a thousand zodis of dust to the inner few AU if the stellar mass loss rate were greater than 100 times solar.

Dust from a more distant source must migrate over a larger distance to reach the inner disk and this migration must be completed in an amount of time roughly equal to the grain-grain collision time.  This means that for a given collision time in a disk, the dominant dust grains that arrive from a more distant source must migrate more quickly, i.e. all other things being equal, more distant sources will deliver smaller dust grains to the inner disk.  Since planets with $M_p<$ M$_{\rm N}$ don't trap the smallest dust grains with 100\% efficiency, we expect the amplitude of the dust transit signal to decrease for these models.  More massive planets still trap small grains with very high efficiency, though, so we expect our $M_p=1$ M$_{\rm J}$ to remain relatively unchanged in amplitude for more distant sources delivering equal amounts of dust to the inner disk.  However, the shape of the transit signal may change slightly, since small grains exhibit a stronger asymmetry reversal in our models (see Section \ref{asymmetry_section}).

Our models did not consider scenarios in which the planetesimals themselves were trapped in external MMRs.  In such a scenario, the largest grains generated by planetesimal collisions would remain in resonance, while the smaller grains would feel the effects of radiation pressure and immediately occupy orbits with larger semi-major axes and different eccentricities.  Depending on the collision rate of the small grains, PR drag may act quickly enough to bring these small grains back into resonance, possibly creating resonant structures with even larger transit signals than we modeled.  The lifetime of a resonant planetesimal belt would be even shorter than the lifetime of our modeled dynamically cold planetesimal belt, though, since the mean motion resonances would enhance the collision rate of the planetesimals \citep{sk09}, eroding the disk at a faster rate.

\subsection{Photometric detection \& sources of confusion}

\emph{Kepler} was designed to detect the transit signature of an Earth-sized planet, requiring precise photometry on the order of one part in 10 million, but only over timescales of a few hours.  By modeling and removing the long-term systematic variations of \emph{Kepler}, phase-folding a light curve obtained over many orbits, and binning the data to increase the signal to noise ratio, one can improve the long-term photometric precision.  For example, \citet{wos10} detected the exoplanet-induced ellipsoidal stellar variability of Hat-P-7, with a signal amplitude $\sim 10^{-4}$, after 33.5 days of observation, or 15 orbits of the perturbing exoplanet Hat-P-7b.

Our models of exozodiacal resonant ring structures have orbital periods much longer than that of Hat-P-7b; a planet orbiting at 0.1 AU from a Sun-like star takes 11.6 days to complete an orbit.  Since the dust sublimation radius prohibits us from considering planets on smaller semi-major axes, the detection of an exozodiacal resonant ring structure around a Sun-like star with \emph{Kepler} requires a deep understanding and precise modeling of \emph{Kepler}'s long-term systematic variability.  Such modeling is beyond the scope of this paper, as the \emph{Kepler} Pre-search Data Conditioning (PDC) software has not been made public.

In addition to modeling the long-term stability of \emph{Kepler}, future efforts to uncover the transit signal of an exozodiacal resonant ring within the \emph{Kepler} data will need to contend with any possible sources of confusion.  For example, in rare circumstances stellar variability from star spots may resemble our exozodi transit signals if the stellar rotation period is close to one half or any integer multiple of the planet's orbital period and the distribution of star spots creates two minima that lead and trail the planet.  Although there is evidence for the synchronization between a star's rotational period and a Jupiter-mass planet's orbital period near 0.1 AU \citep{lam09}, differential rotation of the star spots as a function of stellar latitude and their relatively short lifetimes make this scenario short-lived and therefore distinguishable from the fixed-period exozodi signal.

Additional planets may also affect observational data to create seemingly broad transit curves under special circumstances.  For a system with two planets in the $2\,:\,1$ MMR, the phase folded light curve of the inner planet would also show the transit of the exterior planet.  The exterior planet's transit would consistently occur at roughly the same phase of the inner planet and would not produce a broad transit signal.  But \citet{lrf11} showed that many multiple planet candidate systems detected by \emph{Kepler} exhibit planet candidates that are \emph{near}, but not within, mean motion resonant configurations.  For such a system, the phase folded light curve of an inner planet would also show the transit signatures of all exterior planets near $j$\,:\,1 resonances, but those transits would drift in planetary phase.  This drift would broaden the transit of the exterior planet by
\begin{equation}
	\delta \phi = 2 \pi \frac{t_{\rm obs}}{T_o} \left| j - \frac{T_o}{T_i} \right|,
\end{equation}
in the limit $|jT_i - T_o| \ll T_i \ll t_{\rm obs}$, where $t_{\rm obs}$ is the total observation time, $T_i$ is the orbital period of the inner planet selected for phase-folding, and $T_o$ is the orbital period of any exterior planet near a $j$\,:\,1 mean motion resonance.  If two roughly terrestrial-sized exterior planets near $j$\,:\,1 resonances transit $\sim 90^{\circ}$ before and after the inner planet's transit, the composite broadened transit signatures may look similar to our resonant ring transit.  A planet near an interior 2\,:\,1 mean motion resonance could also contribute a broadened transit curve that occurs twice per folded orbit, similar to our resonant ring transit curve.  In these cases, a simple periodogram and the non-phase folded light curve should distinguish between additional planets and a resonant ring structure.

As Figure \ref{lightcurve_figure2} shows, the transit curves of resonant rings exhibit several features that may aid in their disambiguation, including minima that may be separated by more than 180$^{\circ}$ in planetary phase and a broad maximum near a planetary phase of 180$^{\circ}$ that peaks at a value less than the overall maximum.  However, low signal to noise may make identifying these features difficult.  Future missions with better photometric precision than \emph{Kepler} over longer time scales would facilitate the detection of these disk features.  Simple two- or three-band transit photometry may also help identify the non-gray transit features expected for small dust grains, e.g. the predicted color asymmetry between the leading and trailing clumps as discussed in Section \ref{asymmetry_section}.

Because resonantly trapped dust may originate from parent bodies with a distribution of inclinations up to a few tens of degrees, the transit signature of a resonant ring structure may be detectable over a slightly wider range of system inclinations than planetary transits.  Identifying the signal unambiguously without any accompanying planetary data may be a difficult task for \emph{Kepler}.  Detection of the weak photometric variations of the disk ($\sim10^{-5}$) over orbital time scales will require multiple rotations of the ring (i.e. phase-folding) and broad time binning to achieve adequate SNR.  Prior knowledge of the planet’s orbital period provides us with the proper phase-folding period to perform this task.  Additionally, the phase between the disk and planetary transits provides the primary evidence linking the observed signal to a resonant ring of dust produced by a planet.

Detection of disk transit signals with \emph{Kepler} may open up a new region of parameter space for exozodi detection.  \emph{Spitzer} was sensitive only to dense exozodis ($\sim1400$ zodis) within 1 AU of nearby solar-type stars \citep{b06}.  Within the decade, the James Webb Space Telescope may extend this detection limit to stars as distant as the typical \emph{Kepler} field star, assuming similar spectral calibration uncertainties.  Within the next few years, the Large Binocular Telescope Interferometer may detect exozodiacal disks as tenuous as 10 zodis, but only for nearby stars \citep{ltu09}.  In the immediate term, \emph{Kepler} observations of disk transits provide the only method for probing asymmetric structure in exozodiacal clouds around distant stars, possibly enabling the detection of exozodiacal dust disks with densities $\sim100$ zodis or less.

\subsection{Caveats \label{caveats_section}}

The signal amplitudes shown in Figure \ref{amp_vs_zodis_figure} assume blackbody dust grains ($Q_{\rm abs}=1$) with an albedo of $\omega=0.5$.  Dust in the zodiacal cloud has an albedo significantly lower, $\sim0.1$ \citep{dlr88, hzc02}, while recent observations of the outer regions of debris disks reveal dust albedos ranging from $\sim0.05$ \citep{k10, kgc05} to $>0.4$ \citep{wbs99,dws08}.  As long as our assumption of $Q_{\rm abs}=1$ is approximately valid at visible wavelengths, an albedo of 0 would only reduce the modeled signal by a factor of 2.

As previously mentioned, our treatment of grain-grain collisions ignores grain fragmentation.  We expect fragments with $\beta > 0.5$ to affect the results only marginally, since these grains leave the system on dynamical timescales much shorter than the collision time.  To first order, including fragments with $\beta < 0.5$ would increase the number of small grains in the inner disk and decrease the number of large grains, since the additional fragments would enhance the collision rate.  As a result, massive planets that readily trap small grains ($M_p \sim 1$ M$_{\rm J}$) may produce an even more pronounced ring structure, while planets that do not trap such small grains should exhibit a weaker ring structure.

The effects of fragments with $\beta \approx 0.5$ is less clear.  \citet{sc06} showed that this population of highly eccentric, barely-bound grains is unusually long-lived compared with other grains and dominates a disk's optical depth exterior to its birth ring.  Production of these smallest fragments in the inner disk likely produces results similar to the $\beta < 0.5$ fragments, but their near-unity eccentricities would make resonant trapping less likely.  Future work to improve grain fragmentation in the collisional grooming algorithm will shed light on these issues.

Perhaps the largest caveat of our models is that we assume a single planet on a circular orbit co-planar with the disk.  A planet with non-zero eccentricity or inclination will trap dust less efficiently than its circular, un-inclined counterpart and/or form resonant structures with less asymmetry.  For example, \citet{dm05} showed that Jupiter-mass planets orbiting at 45 AU with eccentricity $\lesssim 0.1$  trap dust into clumpy resonant structures similar to that shown in the top-left panel of Figure \ref{disk_od_figure}, but the same planet with an orbital eccentricity equal to 0.3 produces a ring without clumps.  Massive planets on eccentric orbits may also create resonant dust rings that rotate at half the rate of the planet's orbital frequency, requiring two orbits of the planet to observe both exozodi transit minima \citep{kh03}.

Additional planets may also reduce the population of dust in a planet's resonances.  Gravitational perturbations by nearby planets could reduce a planet's dust trapping efficiency or reduce the amount of time dust spends in resonance.  Additional planets located between the inner planet and the planetesimal belt could also eject in-spiraling dust grains before they reach the inner planet's MMRs.  Jupiter and Saturn shield the inner few AU of our Solar System, ejecting more than 80\% of in-spiraling Kuiper Belt dust \citep{lzd96,mmm03,ks10}.  Planetary system architectures such as this would greatly reduce the likelihood of a massive exozodiacal cloud being fed by a distant Kuiper Belt-like source of dust.

Our models also assume that 100\% of the dust is generated by a dynamically cold population of planetesimals.  Dust generated by a dynamically cold population of planetesimals, like the asteroid belt, has a higher chance of being resonantly trapped by a planet when compared to dust from a dynamically hot population, like an inclined, eccentric population of comets.  In the zodiacal cloud, the relative amounts of dust at 1 AU supplied by asteroids and comets is unknown; estimates of the relative dust contribution from the asteroid belt range widely from $\sim5$\% to $\sim50$\% \citep{ikm08, njl10}.  However, we predict that exozodiacal transit signals will only be detectable in the near future in systems that are by definition much different from the Solar System, with Jupiter-mass planets orbiting interior to the orbit of Mercury.  We should not expect the planetesimal distributions in these systems to mirror our own.

\section{Conclusions}

We used a collisional grooming algorithm to model the dust distributions of collisional exozodiacal clouds perturbed by single planets on circular orbits around Sun-like stars.  We imaged these dust disks edge-on at visible wavelengths as a function of planetary phase to synthesize transit light curves of structured exozodiacal clouds.  The ``clumps" in an exozodiacal cloud, formed by the planet's resonant trapping of in-spiraling dust, occult starlight and typically create two broad transit minima that lead and trail the planetary transit.  

We measured the transit depth of these light curves as a function of planet mass and semi-major axis, disk density, and the dust grains' optical scattering parameters.  We found that Jupiter-mass planets produce the largest transit signals for exozodi structures, with amplitudes up to $\sim 10^{-4}$.  Neptune-mass planets and super-Earths can create exozodi transit amplitudes $\sim 10^{-5}$ and $\sim 10^{-6}$, respectively, while Earth-mass planets produce weak resonant ring structures for zodi levels $\gtrsim10$ zodis.  Planets more massive than Jupiter produce exozodi transit signals with amplitudes less than $10^{-4}$.

Disks with resonant ring structures possibly detectable with \emph{Kepler} require optical depths approaching $\sim$100 zodis.  Compact disks this massive could have lifetimes of 100 Myr or longer, consistent with a non-transient scenario.  The dust could also originate from more distant sources and migrate into the inner regions of disks.

For a Sun-like star, exozodi transit signals vary on timescales greater than a few days.  Detection of an exozodi transit signal from \emph{Kepler} data will require accurate modeling of \emph{Kepler}'s long-term systematic variability as well as any stellar variability.  Future multi-band transit photometry with improved precision over timescales of many weeks would aid in the detection of the transit signatures of resonant exozodi structures.

Asymmetric structure in exozodiacal clouds may be a critical source of astrophysical noise for future exo-Earth imaging missions \citep{dah09, ltu09}.  This new method of resonant ring transit detection provides the only current probe by which we can place upper limits on the degree of exozodi structure.  This method may also open up a new region of parameter space for exozodi detection around distant stars.

\acknowledgments

The author thanks Alycia Weinberger, Marc Kuchner, Evgenya Shkolnik, Ruth Murray-Clay, John Debes, and Brian Jackson for helpful discussions.  This work was supported by the Carnegie Institution of Washington.


\begin{thebibliography}{}


\bibitem[Absil et al.(2006)]{adm06}Absil, O., et al. 2006, A\&A, 452, 237

\bibitem[Absil et al.(2009)]{aml09}Absil, O., Mennesson, B., Le Bouquin, J.-B., Di Folco, E., Kervella, P., \& Augereau, J.-C. 2009, \apj, 704, 150

\bibitem[Akeson et al.(2009)]{acm09}Akeson, R. L., et al. 2009, \apj, 691, 1896

\bibitem[Augereau et al.(1999)]{alm99}Augereau, J. C., Lagrange, A. M., Mouillet, D., Papaloizou, J. C. B., \& Grorod, P. A. 1999, A\&A, 348, 557

\bibitem[Barnes \& Fortney(2004)]{bf04}Barnes, J. W. \& Fortney, J. J. 2004, \apj, 616, 1193

\bibitem[Beichman et al.(2005)]{b05}Beichman, C. A., et al. 2005, \apj, 626, 1061

\bibitem[Beichman et al.(2006)]{b06}Beichman, C. A., et al. 2006, \apj, 639, 1166

\bibitem[Bloemen et al.(2011)]{bmo11}Bloemen, S., et al. 2011, MNRAS, 410, 1787

\bibitem[Borucki et al.(2010)]{bkb10}Borucki, W. J., et al. 2010, Science, 327, 977

\bibitem[Burns et al.(1979)]{bls79}Burns, J. A., Lamy, P. L., \& Soter, S. 1979, Icarus, 40, 1

\bibitem[Chiang et al.(2009)]{ckk09}Chiang, E., Kite, E., Kalas, P., Graham, J. R., \& Clampin, M. 2009, \apj, 693, 734

\bibitem[Chiang \& Jordan(2002)]{cj02}Chiang, E. I., \& Jordan, A. B. 2002, \aj, 124, 3430

\bibitem[Debes et al.(2008)]{dws08}Debes, J. H., Weinberger, A. J., \& Schneider, G. 2008, ApJ, 673, L191

\bibitem[Defr\`ere et al.(2009)]{dah09}Defr\`ere, D., Absil, O., den Hartog, R., Hanot, C., \& Stark, C. 2009, A\&A, 509, 9

\bibitem[Deller \& Maddison(2005)]{dm05}Deller, A. T., \& Maddison, S. T. 2005, ApJ, 625, 398

\bibitem[Dermott et al.(1994)]{d94}Dermott, S. F., Jayaraman, S., Xu, Y. L., Gustafson, B. \AA. S., \& Liou, J. C. 1994, Nature, 369, 719

\bibitem[Dohnanyi(1969)]{d69}Dohnanyi, J. S. 1969, J. Geophysical Research, 74, 2531

\bibitem[Dumont \& Levasseur-Regourd(1988)]{dlr88}Dumont, R. \& Levasseur-Regourd, A. C.,  1988, A\&A, 191, 154

\bibitem[Gilliland et al.(2010)]{gjb10}Gilliland, R. L., et al. 2010, \apj, 713, L160

\bibitem[Golimowski et al(2006)]{g06}Golimowski, D. A., et al. 2006, ApJ, 131, 3109

\bibitem[Greaves et al.(1998)]{g98}Greaves, J. S., et al. 1998, ApJ, 506, 133

\bibitem[Greaves et al.(2005)]{ghw05}Greaves, J. S., et al. 2005, \apj, 619, L187

\bibitem[Hahn et al.(2002)]{hzc02}Hahn, J. M., Zook, H. A., Cooper, B., \& Sunkara, B. 2002, Icarus, 158, 360

\bibitem[Heap et al.(2000)]{h00}Heap, S. R., Lindler, D. J., Lanz, T. M., Cornett, R. H., Hubeny, I., Maran, S. P., \& Woodgate, B. 2000, ApJ, 539, 435

\bibitem[Henyey \& Greenstein(1941)]{hg41}Henyey, L. G., \& and Greenstein, J. L. 1941, \apj, 93, 70

\bibitem[Hong(1985)]{h85}Hong, S. S. 1985, A\&A, 146, 67

\bibitem[Ipatov et al.(2008)]{ikm08}Ipatov, S. I., Kutyrev, A. S., Madsen, G. J., Mather, J. C., Moseley, S. H., \& Reynolds, R. J. 2008, Icarus, 194, 769

\bibitem[Jackson \& Zook(1989)]{jz89}Jackson, A. A., \& Zook, H. A. 1989, Nature, 337, 629

\bibitem[Jenkins et al.(2010)]{jcc10}Jenkins, J. M., et al. 2010, \apj, 713, L120

\bibitem[Kalas et al.(2005)]{kgc05}Kalas, P., Graham, J. R., \& Clampin, M. 2005, Nature, 435, 1067

\bibitem[Kelsall et al.(1998)]{k98}Kelsall, T., et al. 1998, ApJ, 508, 44

\bibitem[Koch et al.(2010)]{kbb10}Koch, D. G., et al. 2010, \apj, 713, L79

\bibitem[Krist et al.(2010)]{k10}Krist, J. E., et al. 2010, AJ, 140, 1051

\bibitem[Krist et al.(2005)]{kag05}Krist, J. E., et al. 2005, \aj, 129, 1008

\bibitem[Krivov et al.(2006)]{kls06}Krivov, A.V., L\"{o}hne, T., \& Srem\u{c}evi\'c, M. 2006, A\&A, 455, 509

\bibitem[Kuchner \& Stark(2010)]{ks10}Kuchner, M. J., \& Stark, C. C. 2010, \aj, 140, 1007

\bibitem[Kuchner \& Holman(2003)]{kh03}Kuchner, M. J., \& Holman, M. J. 2003, \apj, 588, 1110

\bibitem[Lanza et al.(2009)]{lam09}Lanza, A. F., et al. 2009, A\&A, 506, 255L

\bibitem[Lawson et al.(2009)]{ltu09}Lawson, P. R., Traub, W. A., \& Unwin, S. C. 2009, Exoplanet Community Report (JPL Publication 2009-3)

\bibitem[Liou et al.(1996)]{lzd96}Liou, J.-C., Zook, H. A., \& Dermott, S. F. 1996, Icarus, 124, 429

\bibitem[Lissauer et al.(2011)]{lrf11}Lissauer, J. J., et al. 2011, arXiv:1102.0543v3

\bibitem[Melis et al.(2010)]{mzr10}Melis, C., Zuckerman, B., Rhee, J. H., \& Song, I. 2010, \apj, 717, L57

\bibitem[Millan-Gabet et al.(2011)]{mgsm11}Millan-Gabet, R., et al. 2011, \apj, 734, 67

\bibitem[Mo\'or et al.(2009)]{map09}Mo\'or, A., et al. 2009, \apj, 700, L25

\bibitem[Moro-Mart\'{i}n \& Malhotra(2003)]{mmm03}Moro-Mart\'{i}n, A., \& Malhotra, R. 2003, AJ, 125, 2255

\bibitem[Murray-Clay \& Chiang(2005)]{mcc05}Murray-Clay, R. A., \& Chiang, E. I. 2005, \apj, 619, 623

\bibitem[Mustill \& Wyatt(2011)]{mw11}Mustill, A. J., \& Wyatt, M. C. 2011, MNRAS, 413, 554

\bibitem[Nesvorn\'{y} et al.(2010)]{njl10}Nesvorn\'{y}, D., Jenniskens, P., Levison, H. F., Bottke, W. F., Vokrouhlick\'{y}, D., \& Gounelle, M. 2010, ApJ, 713, 816

\bibitem[Quillen(2006)]{q06}Quillen, A. C. 2006, MNRAS, 365, 1367

\bibitem[Reach et al.(1995)]{r95}Reach, W. T., et al. 1995, Nature, 374, 521

\bibitem[Rhee et al.(2008)]{rsz08}Rhee, J. H., Song, I., \& Zuckerman, B. 2008, \apj, 675, 777

\bibitem[Sartoretti, \& Schneider(1999)]{ss99}Sartoretti, P., \& Schneider, J. 1999, A\&AS, 134, 553

\bibitem[Schneider et al.(2009)]{s09}Schneider, G., Weinberger, A. J., Becklin, E. E., Debes, J. H., Smith, B. A. 2009, \aj, 137, 53

\bibitem[Seager et al.(2007)]{skh07}Seager, S., Kuchner, M., Hier-Majumder, C. A., \& Militzer, B. 2007, \apj, 669, 1279

\bibitem[Song et al.(2005)]{szw05}Song, I., Zuckerman, B., Weinberger, A. J., \& Becklin, E. E. 2005, Nature, 436, 363

\bibitem[Stark \& Kuchner(2008)]{sk08}Stark, C. C., \& Kuchner, M. J. 2008, \apj, 686, 637

\bibitem[Stark \& Kuchner(2009)]{sk09}Stark, C. C., \& Kuchner, M. J. 2009, \apj, 707, 543

\bibitem[Strubbe \& Chiang(2006)]{sc06}Strubbe, L. E., \& Chiang, E. I. 2006, \apj, 648, 652

\bibitem[Vitense et al.(2010)]{vkl10}Vitense, C., Krivov, A. V., \& L\"{o}hne, T. 2010, A\&A, 520, 32

\bibitem[Weinberger et al.(1999)]{wbs99}Weinberger, A.J., Becklin, E.E., Schneider, G., Smith, B.A., Lowrance, P.J., Silverstone, M.D., Zuckerman, B., \& Terrile, R.J., 1999, ApJ, 525, 53

\bibitem[Weinberger et al.(2011)]{wbs11}Weinberger, A. J., Becklin, E. E., Song, I., \& Zuckerman, B. 2005, \apj, 726, 72

\bibitem[Welsh et al.(2010)]{wos10}Welsh, W. F., Orosz, J. A., Seager, S., Fortney, J. J., Jenkins, J., Rowe, J. F., Koch, D., \& Borucki, W. J. 2010, \apj, 713L, 145

\bibitem[Wisdom(1980)]{w80}Wisdom, J. 1980, \aj 85, 1122

\bibitem[Wyatt(2003)]{w03}Wyatt, M. C. 2003, \apj, 598, 1321

\bibitem[Wyatt(2006)]{w06}Wyatt, M. C. 2006, \apj, 639, 1153

\bibitem[Wyatt et al.(2005)]{wgd05}Wyatt, M. C., Greaves, J. S., Dent, W. R. F., \& Coulson, I. M. 2005, \apj, 620, 492

\bibitem[Wyatt et al.(2007)]{wsg07}Wyatt, M. C., Smith, R., Greaves, J. S., Beichman, C. A., Bryden, G., \& Lisse, C. M. 2007, \apj, 658, 569

\end{thebibliography}
\end{document}